\documentclass[%
 reprint,
 amsmath,amssymb,
 aps,
]{revtex4-2}

\usepackage{graphicx}
\usepackage{dcolumn}
\usepackage{bm}

\begin{document}

\preprint{APS/123-QED}

\title{Quantum theory of cross-correlation heterodyne detection}
\thanks{A footnote to the article title}%

\author{Sheng Feng}
 \email{fensf2a@hust.edu.cn}
\author{Kaikai Wu}
\altaffiliation[Also at ]{Hubei Key Laboratory of Modern Manufacturing Quantity Engineering, School of Mechanical Engineering, Hubei University of Technology, Wuhan, Hubei 430068, China.}
\affiliation{
 School of Electrical and Electronic Infomation Engineering, Hubei Polytechnic University, Huangshi, Hubei 435003, P.R. China
}%


\date{\today}

\begin{abstract}
Cross-correlation heterodyne detectors exhibit the potential for suppression of the detection quantum noise below shot noise without use of optical squeezing for capturing weak optical signals in low frequency bands. To
understand the underlying mechanism, we develop a quantum theory to describe the noise performance of cross-correlation heterodyne detectors. By calculating the cross spectral density (CSD) of the photocurrent fluctuations
from a cross-correlation heterodyne detector, we prove that its noise performance can break the shot noise limit and exceed that of a regular heterodyne detector for detection of coherent light. When the detected light signal
is in a squeezed state, we show that the corresponding CSD value is negative and discuss how a negative CSD may be explored to improve the output signal-to-noise ratio of the detector contaminated by classical noises through
tuning the parameter of the degree of squeezing. This work may find itself useful in space-based gravitational wave searching and a variety of other scientific research activities, such as observation of vacuum magnetic
birefringence and telecommunications.
\end{abstract}

\maketitle


\section{Introduction}
Heterodyne detectors are powerful tools for capturing low-frequency weak optical signals and exclusively exploited in space-borne gravitational wave (GW) astronomy \cite{armano2006,luo2016,babak2017}, vacuum magnetic
birefringence observation \cite{zavattini2006} and telecommunications. As the space-based GW observation systems are approaching to their quantum noise limits \cite{Sesana2016}, the quantum property of heterodyne detectors
will become a limiting factor to further improvement of GW detection sensitivity. Although squeezed states of light may be utilized to reduce the quantum noise in heterodyne detection of audio-band weak optical signals
\cite{xie2018}, photon loss in space-based GW observation systems will substantially deteriorate the squeezing enhancement of the signal-to-noise ratio (SNR) in GW searching. Recently, cross-correlation heterodyne detection
has been demonstrated for potential application in the suppression of the detection quantum noise below shot noise \cite{michael2018}. Yet a quantum theory is highly desired for a full understand of the quantum noise behavior
of cross-correlation heterodyne detectors.

\begin{figure}[htbp]
\centering
\includegraphics[width=7cm]{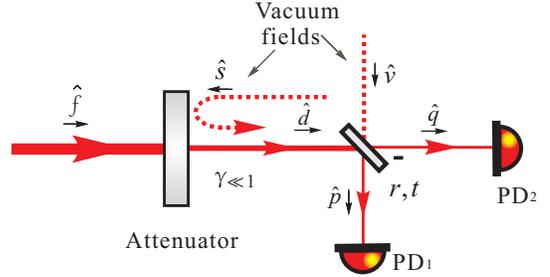}
\caption{Illustration of cross-correlation direct detection below shot noise.} \label{fig:ccdd}
\end{figure}

Cross correlation detection, which is also referred to as quantum correlation measurement, has been extensively investigated in the configuration of direct detection or homodyne detection
\cite{mitsui2013,martynov2017,venneberg2022}. The idea of cross-correlation direct detection of an attenuated optical signal below shot noise is illustrated in Fig. \ref{fig:ccdd} \cite{venneberg2022}. The original signal
field $\hat{f}$ to be detected undergoes optical attenuation, for whatever reasons, and serious SNR degradation will be caused in usual detection. It has been shown that a vacuum field, say $\hat{s}$, is introduced into the
beam during the signal attenuation and this SNR degradation can be overcome in cross-correlation detection \cite{venneberg2022}. To this end, the attenuated signal field $\hat{d}$ is split into two parts by a beamsplitter
(amplitude reflectivity $r$ and transmissivity $t$) where another vacuum field $\hat{v}$ is mixed with the field $\hat{d}$. Each of the output fields ($\hat{p}$ and $\hat{q}$) from the beamsplitter is received by a
photodetector PD$_1$ or PD$_2$.

It can be shown that the output fields, $\hat{p}$ and $\hat{q}$, are related to the original signal field and vacuum fields by \cite{venneberg2022}
\begin{eqnarray} \label{eq:pq}
\hat{p}&=& r \gamma \hat{f} + r \sqrt{1-\gamma^2}\hat{s}+t\hat{v}, \nonumber \\
\hat{q}&=& t \gamma \hat{f} + t \sqrt{1-\gamma^2}\hat{s}-r\hat{v}\ ,
\end{eqnarray}
wherein $\gamma$ denotes the amplitude attenuation coefficient of the original signal field. The cross-correlation direct detection is carried out by computing the cross spectral density (CSD) $S_{ii}^{\hat{p} \hat{q}}$ of two
field quadratures $\hat{p}_i$ and $\hat{q}_i$ ($i=1,2$), \cite{venneberg2022}
\begin{eqnarray} \label{eq:csdpq}
S_{ii}^{\hat{p} \hat{q}}&=& rt\left[\gamma^2 S_{ii}^{\hat{f}} + (1-\gamma^2)S_{ii}^{\hat{s}}-S_{ii}^{\hat{v}}\right]\ ,
\end{eqnarray}
in which $S_{ii}^{\hat{f}}, S_{ii}^{\hat{s}}, S_{ii}^{\hat{v}}$ are the auto (power) spectral densities of the corresponding field quadratures $\hat{f}_i, \hat{s}_i, \mbox{or} \ \hat{v}_i$, respectively. Since $\hat{s}$ and
$\hat{v}$ are in vacuum states, one has $S_{ii}^{\hat{s}}=S_{ii}^{\hat{v}}=1$. Then it follows that
\begin{eqnarray} \label{eq:ssnl}
S_{ii}^{\hat{p} \hat{q}}&=& rt\gamma^2(S_{ii}^{\hat{f}}-1)\ .
\end{eqnarray}
If the original signal field $\hat{f}$ is in a coherent state with negligible classical noises, then $S_{ii}^{\hat{f}}\approx1$, which means that $S_{ii}^{\hat{p} \hat{q}}\approx 0$ following Eq. (\ref{eq:ssnl}). In other
words, cross-correlation direct detection allows one to achieve a reduction of detection quantum noise below the shot noise level in the detection of coherent light.


Nonetheless, the above theoretical analysis cannot be naively applied to the case of heterodyne detection that turns out to be more complex than direct/homodyne detection. We start from the quantum theory of optical coherence
\cite{glauber1963} and develop in this work a theoretical description of cross-correlation heterodyne detection. In the next section, we will establish a theoretical model describing the output signal and the CSD of the
photocurrent fluctuations produced by the detector as a Fourier transform of a two-time cross-correlation function. Then in Section 3 we will consider cross-correlation heterodyne detection of squeezed light that is created
through some nonlinear interaction process such as parametric down-conversion or four-wave mixing. One will see that the value of the CSD is negative when squeezed light is fed into the detector, which is mathematically
allowed though \cite{mandel1995}. Finally, the fourth section is devoted into discussions about the reduction of the detection quantum noise below shot noise in cross-correlation heterodyne detection of coherent light, the
meaning of negative CSD and how it affects the detection noise levels in cross-correlation heterodyne detection of squeezed light.

\section{Theoretical model}
To begin with, let consider a quantum field of signal light with a continuum of frequency modes \cite{feng2016,glauber1963,mandel1995},
\begin{equation}\label{eq:field}
\hat{E}_{\rm s}^{(+)}(\mathbf{r},t)=\frac{\rm i}{\sqrt{\varepsilon_0V}} \sum_{\mathbf{k}}\left(\frac{1}{2}\hbar\omega_{\mathbf{k}}\right)^{\frac{1}{2}} \hat{a}_{\mathbf{k}}{\rm e}^{{\rm
i}(\mathbf{k}\cdot\mathbf{r}-\omega_{\mathbf{k}}t)},
\end{equation}
where $V$ is the quantization volume, $\varepsilon_0$ represents the dielectric permittivity of vacuum, $\mathbf{k}$ denotes a set of plane-wave modes with $\omega_{\mathbf{k}}$ the corresponding angular frequency of each
mode, and $\hbar\equiv h/2\pi$ in which $h$ is Planck constant. The amplitude operator $\hat{a}_{\mathbf{k}}$ stands for the photon annihilation operator for mode $\mathbf{k}$ and stays constant if there is no free electrical
charge in the space \cite{glauber1963}. The two mutually adjoint operators $\hat{a}_{\mathbf{k}}$ and $\hat{a}^\dagger_{\mathbf{k}}$ obey the following commutators,
\begin{equation}
\left[\hat{a}_{\mathbf{k}}, \hat{a}_{\mathbf{k'}}\right]=\left[\hat{a}^\dagger_{\mathbf{k}}, \hat{a}^\dagger_{\mathbf{k'}}\right] = 0,\ \left[\hat{a}_{\mathbf{k}},
\hat{a}^\dagger_{\mathbf{k'}}\right]=\delta_{\mathbf{k},\mathbf{k'}}.
\end{equation}

\begin{figure}[htbp]
\centering
\includegraphics[width=8cm]{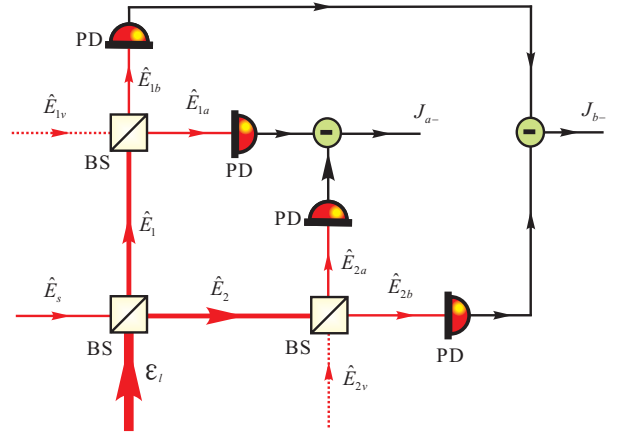}
\caption{Schematic of cross-correlation heterodyne detection of optical signals. BS: 50-50 beamsplitter. PD: Photodiode.} \label{fig:model}
\end{figure}

Suppose that the signal field is fed into a heterodyne detector of  conventional balanced configuration, except that each of the two output beams from the first beamsplitter are farther equally split into two parts resulting
in four output optical beams in total (Fig. \ref{fig:model}). When each of the light beam is received by a photodiode, it generates photoelectric emissions at certain times $t_1, t_2, ...$. If $j(t)$ is the current output
pulse created by a photoemission at time $t_i=0$, then the total photoelectric current $J(t)$ can be represented by the sum of pulses over all times $t_i$,
\begin{equation}
J(t)=\sum_i j(t-t_i).
\end{equation}
Apparently, $j(t-t_i)=0$ if $t<t_i$ and the summation may be converted into integral in the following calculations.

The four output light beams are paired as depicted in Fig. \ref{fig:model} and each pair of beams produces one differential photocurrent, $J_{\rm a-}(t)$ or $J_{\rm b-}(t)$, \cite{feng2016,ou1987}
\begin{eqnarray}\label{eq:jminus}
&&J_{\rm a-}(t)=\eta \int^{\infty}_0 {\rm d} t' j(t')<\hat{I}_{\rm 2a}(t-t')-\hat{I}_{\rm 1a}(t-t')> \nonumber\\
&&J_{\rm b-}(t)=\eta \int^{\infty}_0 {\rm d} t' j(t')<\hat{I}_{\rm 2b}(t-t')-\hat{I}_{\rm 1b}(t-t')> ,
\end{eqnarray}
where $\eta$ is the quantum efficiency of the detector in units of the average number of photoelectrons per photon energy, $\hat{I}(t)\equiv c \varepsilon_0\hat{E}^{(-)}(t)\hat{E}^{(+)}(t)$ is the optical intensity sensed by a
photodiode,  $\hat{E}^{(-)}(t)=[\hat{E}^{(+)}(t)]^\dagger$, $c$ stands for the speed of light in vacuum,  and $<\cdot>$ denotes statistical averaging. Using the operation matrix for a 50-50 beamsplitter,
\begin{eqnarray}
\frac{1}{\sqrt{2}}\left[
\begin{array}{cc}
1 & i\\
i & 1
\end{array}
\right],
\end{eqnarray}
one may obtain the output fields,
\begin{eqnarray} \label{eq:outfields}
&&\hat{E}^{(+)}_{\rm 1a}(t)=[\hat{E}^{(+)}_{\rm s}(t)+\sqrt{2}i\hat{E}^{(+)}_{\rm 1v}(t)+i\varepsilon^{(+)}_{\rm l}(t)]/2 \nonumber \\
&&\hat{E}^{(+)}_{\rm 1b}(t)=[i\hat{E}^{(+)}_{\rm s}(t)+\sqrt{2}\hat{E}^{(+)}_{\rm 1v}(t)-\varepsilon^{(+)}_{\rm l}(t)]/2 \nonumber \\
&&\hat{E}^{(+)}_{\rm 2a}(t)=[i\hat{E}^{(+)}_{\rm s}(t)+\sqrt{2}i\hat{E}^{(+)}_{\rm 2v}(t)+\varepsilon^{(+)}_{\rm l}(t)]/2 \nonumber \\
&&\hat{E}^{(+)}_{\rm 2b}(t)=[-\hat{E}^{(+)}_{\rm s}(t)+\sqrt{2}\hat{E}^{(+)}_{\rm 2v}(t)+i\varepsilon^{(+)}_{\rm l}(t)]/2.\
\end{eqnarray}
Let assume fast response speed for the detector such that $j(t)=e\delta(t)$ (where $\delta(t)$ is Dirac function with $e$ being the charge on the electron). From Eqs. (\ref{eq:jminus}) and (\ref{eq:outfields}), the signal
produced in the cross-correlation heterodyne detection reads
\begin{eqnarray}\label{eq:sighet}
S(t)&=&J_{a-}(t)J_{b-}(t)  \nonumber \\
&=& (\eta e c \varepsilon_0)^2 <\hat{I}_{\rm 2a}(t)-\hat{I}_{\rm 1a}(t)>< \hat{I}_{\rm 2b}(t)-\hat{I}_{\rm 1b}(t)> \nonumber \\
&=& C \left[<\varepsilon_{\rm l}^{(+)}(t)\hat{E}^{(-)}_{\rm s}(t)-\varepsilon_{\rm l}^{(-)}(t)\hat{E}^{(+)}_{\rm s}(t)>\right]^2,\  \  \
\end{eqnarray}
in which $C=-(\eta e c \varepsilon_0)^2/4$, the averages of the vacuum fields $\hat{E}^{(+)}_{\rm 1v}(t)$ and $\hat{E}^{(+)}_{\rm 2v}(t)$ are absent since they are zero, $\varepsilon_{\rm l}^{(+)}(t)$ is the local oscillator
(LO) field, and $\varepsilon_{\rm l}^{(-)}(t)=[\varepsilon_{\rm l}^{(+)}(t)]^*$.

Next let turn to the calculation of the CSD of $J_{\rm a-}(t)$ and $J_{\rm b-}(t)$, which is defined as \cite{mandel1995}
\begin{equation}\label{eq:csdhet}
\chi_{csd}(\omega)=\frac{1}{T}{\int}^{T}_{0}{\rm d} t{\int}^{+\infty}_{-\infty}{\rm d}\tau {\rm e}^{{\rm i}\omega \tau}<\Delta J_{\rm a-}(t) \Delta J_{\rm b-}(t+\tau)>,
\end{equation}
wherein $T$ denotes the heterodyne measurement time and $\Delta A\equiv A-<A>$. The cross-correlation function of the differential photocurrent fluctuations is \cite{feng2016,ou1987}
\begin{eqnarray}\label{eq:autoJ}
&&<\Delta J_{\rm a-}(t) \Delta J_{\rm b-}(t+\tau)>\nonumber\\
&=&\sum_{\rm i,\rm j=1}^2 (-1)^{\rm i+\rm j}<\Delta J_{\rm ia}(t) \Delta J_{\rm jb}(t+\tau)>\nonumber \\
&=&\sum_{\rm i,\rm j=1}^2 \eta^2(-1)^{\rm i+\rm j}\times \nonumber\\
&&\int\int_0^{\infty} {\rm d} t'{\rm d} t''j_{\rm ia}(t')j_{\rm jb}(t'') \lambda_{\rm ia;\rm jb}(t-t',\tau+t'-t'').\
\end{eqnarray}
Here $j_{\rm ia,\rm jb}(t)$ are photoemission-induced current pulses at the corresponding photodiodes and the correlation functions of light-intensity fluctuations are \cite{glauber1963}
\begin{eqnarray}\label{eq:lambda}
&&\lambda_{\rm ia;\rm jb}(t,\iota)
=<T:\Delta \hat{I}_{\rm ia}(t) \Delta\hat{I}_{\rm jb}(t+\iota):>\nonumber\\
&=&c^2\varepsilon_0^2< \hat{E}^{(-)}_{\rm ia}(t) \hat{E}^{(-)}_{\rm jb}(t+\iota) \hat{E}^{(+)}_{\rm jb}(t+\iota) \hat{E}^{(+)}_{\rm ia}(t)> \nonumber\\
&-&c^2\varepsilon_0^2< \hat{E}^{(-)}_{\rm ia}(t)  \hat{E}^{(+)}_{\rm ia}(t)><\hat{E}^{(-)}_{\rm jb}(t+\iota) \hat{E}^{(+)}_{\rm jb}(t+\iota)>, \nonumber\\
\end{eqnarray}
wherein the symbol $T: :$  is time- and normal-ordering of the field operators $\hat{E}^{(\pm)}(t)$. Photodiode noises are not included in Eq. (\ref{eq:csdhet}) since at the moment we consider only the situation in which the
system sensitivity in the frequency band of interest is limited by the quantum noise of light \cite{Sesana2016}.

Plugging Eqs. (\ref{eq:outfields}) into Eq. (\ref{eq:lambda}) leads to
\begin{eqnarray}\label{eq:lambdan}
\lambda_{\rm ia;\rm jb}(t,\iota)&=&16^{-1}c^2 \varepsilon_0^2 (-1)^{\rm i+\rm j}\times\nonumber\\
&&\big[<\Delta\hat{E}_{\rm s}^{(-)}(t)\Delta\hat{E}_{\rm s}^{(+)}(t+\iota)>\varepsilon^{(+)}_{\rm l}(t)\varepsilon^{(-)}_{\rm l}(t+\tau) \nonumber \\
&+&<\Delta\hat{E}_{\rm s}^{(-)}(t+\iota)\Delta\hat{E}_{\rm s}^{(+)}(t)>\varepsilon^{(-)}_{\rm l}(t)\varepsilon^{(+)}_{\rm l}(t+\tau)\nonumber \\
&-&<\Delta\hat{E}_{\rm s}^{(-)}(t)\Delta\hat{E}_{\rm s}^{(-)}(t+\iota)>\varepsilon^{(+)}_{\rm l}(t)\varepsilon^{(+)}_{\rm l}(t+\tau) \nonumber \\
&-&<\Delta\hat{E}_{\rm s}^{(+)}(t+\iota)\Delta\hat{E}_{\rm s}^{(+)}(t)>\varepsilon^{(-)}_{\rm l}(t)\varepsilon^{(-)}_{\rm l}(t+\tau)\big] ,\nonumber\\
\end{eqnarray}
from which all the low-order terms in $|\varepsilon_{\rm l}|$ are dropped. In addition, because $\hat{E}_{\rm iv}^{(\pm)}(t)$ (i=1,2) are in vacuum states, all the terms of $<\Delta\hat{E}_{\rm iv}^{(\pm)}(t)\Delta\hat{E}_{\rm
iv}^{(\pm)}(t+\iota)>$ and $<\Delta\hat{E}_{\rm iv}^{(\pm)}(t+\iota)\Delta\hat{E}_{\rm iv}^{(\pm)}(t)>$ are zero \cite{glauber1963} and do not appear in Eq. (\ref{eq:lambdan}).

Let assume a monochromatic LO field, i.e., $\varepsilon_{\rm l}^{(+)}(t)=\varepsilon_{\rm l}e^{-i\omega_{\rm l} t+i\mathbf{k}_{\rm l}\cdot\mathbf{r}+i\theta_{\rm l}}$ ($\varepsilon_{\rm l}, \theta_{\rm l}, \omega_{\rm l},
\mathbf{k}_{\rm l}$ are respectively the amplitude, the phase, the frequency, and the wave vector of the LO field). With the help of the definitions of
\begin{eqnarray}\label{eq:gammad}
&&\Gamma_{\rm s}^{(1,1)}(t,\iota)\equiv <\Delta \hat{E}_{\rm s}^{(-)}(t)\Delta \hat{E}_{\rm s}^{(+)}(t+\iota)>{\rm e}^{{\rm i}\omega_{\rm l}\iota},\nonumber\\
&&\Gamma_{\rm s}^{(2,0)}(t,\iota)\equiv <\Delta \hat{E}_{\rm s}^{(-)}(t)\Delta \hat{E}_{\rm s}^{(-)}(t+\iota)>{\rm e}^{-{\rm i}\omega_{\rm l}(2t+\iota)},\nonumber\\
\end{eqnarray}
Eq. (\ref{eq:lambdan}) may be rewritten as
\begin{eqnarray}\label{eq:app5}
\lambda_{\rm ia;\rm jb}(t,\iota)&=&16^{-1}c^2 \varepsilon_0^2 \varepsilon_{\rm l}^2 (-1)^{\rm i+\rm j}\nonumber\\
&\times& \big[\Gamma_{\rm s}^{(1,1)}(t,\iota)-\Gamma_{\rm s}^{(2,0)}(t,\iota){\rm e}^{2{\rm i}\theta'_{\rm l}}+ {\rm c.c.}\big],
\end{eqnarray}
in which $\theta'_{\rm l}=\mathbf{k}_{\rm l}\cdot\mathbf{r}+\theta_{\rm l}$. Under the approximation of fast response speed for the detector, from Eqs. (\ref{eq:csdhet}), (\ref{eq:autoJ}), and (\ref{eq:app5}) it follows that
\begin{eqnarray}\label{eq:csdhetn}
\chi_{csd}(\omega)&=&2\eta^2 c^2 \varepsilon_0^2 e^2 \varepsilon_{\rm l}^2 \frac{1}{4T}{\int}^{T}_{0}{\rm d} t{\int}^{+\infty}_{-\infty}{\rm d}\tau {\rm e}^{{\rm i}\omega \tau} \times\nonumber \\
&& \big[\Gamma_{\rm s}^{(1,1)}(t,\tau)-\Gamma_{\rm s}^{(2,0)}(t,\tau){\rm e}^{2{\rm i}\theta'_{\rm l}}+ {\rm c.c.}\big],
\end{eqnarray}
where the factor of 2 accounts for the noise contribution of negative-frequency components \cite{feng2016}. Eqs. (\ref{eq:sighet}) and (\ref{eq:csdhetn}) are  generic formulae to describe the quantum noise nature of a
cross-correlation heterodyne detector. In comparison with the case of traditional heterodyne detection, the shot noise term is absent in Eq. (\ref{eq:csdhetn}), i.e., the quantum noise level in  cross-correlation heterodyne
detection can be lower than that of shot noise.

\section{Cross-correlation heterodyne detection of squeezed light}

In this section, we apply the preceding results of Eqs. (\ref{eq:sighet}) and (\ref{eq:csdhetn}) to the detection of squeezed light, which may be produced in different physical processes such as four-wave mixing
\cite{Slusher1985} and parametric down-conversion \cite{{Wu1986}}. The optical field in a squeezed state is related to an input field $\hat{E}_{\rm 0}^{(+)}(t)$ that is usually in a coherent state or a vacuum state,
\begin{equation}\label{eq:0field}
\hat{E}_{\rm 0}^{(+)}(\mathbf{r},t)=\frac{{\rm i}}{\sqrt{\varepsilon_0V}} \sum_{\mathbf{k}}\left(\frac{1}{2}\hbar\omega_{\mathbf{k}}\right)^{\frac{1}{2}} \hat{b}_{\mathbf{k}}{\rm e}^{{\rm
i}(\mathbf{k}\cdot\mathbf{r}-\omega_{\mathbf{k}}t)},
\end{equation}
from which optical squeezing is created, through the linear evolution equation \cite{caves1982},
\begin{equation}\label{eq:evolve}
\hat{a}_{\rm s}=\hat{b}_{\rm s}\cosh r + \hat{b}_{\rm i}^\dagger\sinh r,\hspace{0.1in} \hat{a}_{\rm i}=\hat{b}_{\rm s}^\dagger\sinh r + \hat{b}_{\rm i}\cosh r.
\end{equation}
Here $\hat{b}_{\rm s,\rm i}$ is the photon annihilation operator of the signal/idler mode in $\hat{E}_{\rm 0}^{(+)}(t)$, and $r$ is a real constant determined by the strength and time duration of the squeezing generation.

From Eqs. (\ref{eq:field}) (\ref{eq:0field}), and (\ref{eq:evolve}), it is not difficult to show \cite{xie2021}
\begin{eqnarray}\label{eq:fevolve}
\hat{E}_{\rm s}^{(+)}(\mathbf{r},t)&=&\hat{E}_{\rm 0}^{(+)}(\mathbf{r},t)\cosh r\nonumber\\
&-&\hat{E}_{\rm i}^{(-)}(\mathbf{r},t){\rm e}^{2{\rm i}(\mathbf{k}_{\rm +}\cdot\mathbf{r}-\omega_{\rm +} t)}\sinh r ,
\end{eqnarray}
in which
\begin{equation}\label{eq:ifield}
\hat{E}_{\rm i}^{(+)}(\mathbf{r},t)\equiv\frac{\rm i}{\sqrt{\varepsilon_0V}} \sum_{\mathbf{k}} \left(\frac{1}{2} \hbar |2\omega_{\rm +}-{\omega_{\mathbf{k}}}|\right)^{\frac{1}{2}} \hat{b}_{\mathbf{k}}{\rm e}^{{\rm
i}(\mathbf{k}\cdot\mathbf{r}-\omega_{\mathbf{k}}t)},
\end{equation}
and $\omega_{\rm +}=(\omega_{\rm s}+\omega_{\rm i})/2$, $\mathbf{k}_{\rm +}=(\mathbf{k}_{\rm s}+\mathbf{k}_{\rm i})/2$. Here $\omega_{\rm i}$ and $\mathbf{k}_{\rm i}$ are respectively the frequency and the wave vector of the
idler mode during the generation of optical squeezing.

Now suppose that the optical parameters of the LO field are chosen to satisfy $\omega_{\rm +}=\omega_{\rm l}$ and $\mathbf{k}_{\rm +}=\mathbf{k}_{\rm l}$. From Eqs. (\ref{eq:sighet}) and (\ref{eq:fevolve}) it follows that the
output signal from the detector reads
\begin{eqnarray}\label{eq:sighetn}
S(t)&=& \frac{(\eta e c \varepsilon_0\varepsilon_{\rm l}\beta_{\rm s})^2}{2} \times \nonumber \\
&&\left[\cos\theta_{\rm l}\cos(\Omega t-\theta'_{\rm s})e^r- \sin\theta_{\rm l}\sin(\Omega t-\theta'_{\rm s})e^{-r}\right]^2,\
\end{eqnarray}
wherein $\beta_{\rm s}e^{\theta_{\rm s}}\equiv(\hbar\omega_{\rm s}/\varepsilon_0V)^{1/2}<\hat{b}_{\rm s}>$ is the complex amplitude of the signal mode  of the field $\hat{E}_0^{(+)}(t)$ of Eq. (\ref{eq:0field}),
$\Omega=\omega_{\rm s}-\omega_{\rm l}=\omega_{\rm l}-\omega_{\rm i}$ is the heterodyne frequency, $\theta'_{\rm s} = \Delta\mathbf{k}\cdot\mathbf{r}+\theta_{\rm s}$ ($\Delta \mathbf{k}\equiv \mathbf{k}_{\rm s}-\mathbf{k}_{\rm
l}=\mathbf{k}_{\rm l}-\mathbf{k}_{\rm i}$), and we make use of $\omega_{\rm s}\approx \omega_{\rm i}$ since $|\omega_{\rm s}-\omega_{\rm i}|<<\omega_{\rm s,{\rm i}}$ for heterodyne detection. For detection of squeezed light,
the value of $\theta_{\rm l}$ needs to be controlled to be $\theta_{\rm l}=m\pi+\pi/2$ ($m$ is any integer) and thereby one obtains
\begin{eqnarray}\label{eq:sighetnn}
S(t)= \frac{(\eta e c \varepsilon_0\varepsilon_{\rm l}\beta_{\rm s})^2}{2} \sin^2(\Omega t-\theta'_{\rm s})e^{-2r}.\
\end{eqnarray}

As for the calculation of the corresponding CSD in cross-correlation heterodyne detection of optical squeezing, one may first plug Eq. (\ref{eq:fevolve}) into Eqs. (\ref{eq:gammad}) and obtain \cite{xie2021}
\begin{eqnarray}\label{eq:gammaa}
\Gamma_{\rm s}^{(1,1)}(t,\tau)&=&\sinh^2r \ {\rm e}^{-{\rm i}\omega_{\rm l}\tau}\times \nonumber\\
 &&<\big[\Delta\hat{E}_{\rm i}^{(+)}(t),\Delta\hat{E}_{\rm i}^{(-)}(t+\tau)\big]>\ ,
\end{eqnarray}
and
\begin{eqnarray}\label{eq:gammaaa}
\Gamma_{\rm s}^{(2,0)}(t,\tau)&=&
-\sinh r \cosh r \ \ {\rm e}^{-{\rm i}(\omega_{\rm l}\tau+2\mathbf{k}_{\rm l}\cdot\mathbf{r})}\times \nonumber\\
&& <\big[\Delta\hat{E}_{\rm i}^{(+)}(t),\Delta\hat{E}_{\rm 0}^{(-)}(t+\tau)\big]>\ .
\end{eqnarray}
Albeit $\hat{E}_{\rm 0,\rm i}^{(+)}(t)$ in Eqs. (\ref{eq:0field}) and (\ref{eq:ifield}) are expressed in three dimensional (3D) expansions, all the above calculations are valid for their one dimensional (1D) expansions as
well. For optical fields in the form of collimated beams, one may plug the 1D versions of Eqs. (\ref{eq:0field}) and (\ref{eq:ifield}) into Eqs. (\ref{eq:gammaa}) and (\ref{eq:gammaaa}), resulting in \cite{xie2021}
\begin{eqnarray}\label{eq:gammab}
&&\Gamma_{\rm a}^{(1,1)}(t,\tau)-\Gamma_{\rm a}^{(2,0)}(t,\tau){\rm e}^{2{\rm i}\theta'_{\rm l}}\nonumber\\
&=&\frac{\hbar \sinh r}{2\pi c \varepsilon_0}
{\int}^{+\infty}_{0} {\rm d}\omega' {\rm e}^{{\rm i}(\omega'-\omega_{\rm l}) \tau}\times \nonumber\\
&&\left(\sinh r|2\omega_{\rm l}-\omega'|+{\rm e}^{2{\rm i}\theta_{\rm l}}\cosh r \sqrt{\omega'|2\omega_{\rm l}-\omega'|}\right)\ .
\end{eqnarray}
Here the summation over ${\rm k}$ is replaced by an integral: $(1/V) \sum_{\rm k} \rightarrow (1/2\pi)\int {\rm d} {\rm k}$, in which ${\rm k}=\pm\omega'/c$ is the 1D wave number of the light beam. Substituting Eq.
(\ref{eq:gammab}) into Eq. (\ref{eq:csdhetn}) and after some mathematical manipulations, one may obtain
\begin{eqnarray}\label{eq:csdhetnn}
\chi_{csd}(\omega)=2^{-1} c \varepsilon_0 \eta^2  e^2 \varepsilon_{\rm l}^2\hbar\sinh r F(\omega),
\end{eqnarray}
wherein
\begin{eqnarray}\nonumber
F(\omega)&\equiv&\sinh r \ |\omega_{\rm l}+\omega|+\sinh r \ |\omega_{\rm l}-\omega| \nonumber\\ &&+\left({\rm e}^{2{\rm i}\theta_{\rm l}}\cosh r \sqrt{\omega_{\rm l}^2-\omega^2}+{\rm c.c.}\right)\ .
\end{eqnarray}

Since the LO optical frequency is much higher than the heterodyne frequency, i.e., $\omega_{\rm l}>>\omega$, Eq. (\ref{eq:csdhetnn}) may be rewritten as,
\begin{eqnarray}\label{eq:csdhetnnn}
\chi_{csd}(\omega)=2^{-1} c \varepsilon_0 \eta^2 e^2 \varepsilon_{\rm l}^2\hbar\omega_{\rm l}\left[{\rm e}^{2r}\cos^2\theta_{\rm l}+{\rm e}^{-2r}\sin^2\theta_{\rm l}-1\right].
\end{eqnarray}
From Eq. (\ref{eq:csdhetnnn}) it follows that optical squeezing will reduce the CSD value when the phase of the LO field is chosen appropriately, e.g., $\theta_{\rm l}=m\pi+\pi/2$ ($m$ is any integer) may be the choice. In
this case, the squeezing-affected CSD becomes
\begin{eqnarray}\label{eq:csdhetnnnn}
\chi_{csd}(\omega)=2^{-1} c \varepsilon_0 \eta^2 e^2 \varepsilon_{\rm l}^2\hbar\omega_{\rm l}\left({\rm e}^{-2r}-1\right),
\end{eqnarray}
which is obviously negative since $r$ is a positive real number. However, neither the detected signal power or the detected quantum noise is intuitively negative, which naturally raises an interesting question about how to understand the negative property of the CSD in cross-correlation heterodyne detection to which we will turn soon.

\section{Discussions}
From the results of the theoretical analysis given in the preceding sections, we see that the output signal size in cross-correlation heterodyne detection is smaller than that in regular heteordyne detection, as shown by Eq.
(\ref{eq:sighetnn}). Despite this, cross-correlation heterodyne detection still allows one to achieve SNR enhancement for the output signal.

To see this, let consider a coherent signal at the input of the detector. The calculation results in the last section for squeezed light can be applied to the case of coherent signal by setting $r=0$ for Eqs.
(\ref{eq:sighetnn}) and (\ref{eq:csdhetnnnn}). Thereby, one may have an average value of the detected signal,
\begin{equation}\label{eq:sighetcoh}
\bar{S}_{\rm coh}=\frac{1}{T}\int_0^T {\rm d}t S(t)=(\eta e c \varepsilon_0\varepsilon_{\rm l}\beta_{\rm s})^2/4,\
\end{equation}
and the corresponding CSD,
\begin{eqnarray}\label{eq:csdhetcoh}
\chi_{csd}(\omega)=2^{-1} c \varepsilon_0 \eta^2 e^2 \varepsilon_{\rm l}^2\hbar\omega_{\rm l}\left({\rm e}^{0}-1\right)=0.
\end{eqnarray}
Then it follows that the output signal of (\ref{eq:sighetcoh}) is reduced by a factor of 4 in comparison with the signal measured with a traditional detector \cite{xie2021}, whereas the CSD now becomes zero in the
cross-correlation heterodyne detection, at least in principle; this means that the detected SNR is boosted from a finite value to an ideal value of infinity. Of course an infinite SNR is practically impossible because there
are always residual noises in the detection, including both classical noises and quantum noise due to fluctuations of finite photon number during the measurement time. But the bottom line is that cross-correlation heterodyne
detection does provide a feasible way for sub-shot-noise detection of optical signals with enhanced SNR's, as recently demonstrated in experiment \cite{michael2018}.

An important point revealed by Eqs. (\ref{eq:sighetcoh}) and (\ref{eq:csdhetcoh}) is that the average amplitude of a quantum field, $\hat{E}_{\rm s}^{(+)}(t)$, may be measured as a transfer function of the average amplitudes
of other quantum fields, $\hat{E}_{\rm ia}^{(+)}(t)$ and $\hat{E}_{\rm ib}^{(+)}(t)$ (i=1,2), by means of cross-correlation heterodyne detection, whereas the quantum fluctuations of the measured field $\hat{E}_{\rm
s}^{(+)}(t)$ do not show up in the detection results. This is an essential feature that distinguishes cross-correlation heterodyne detection from regular heterodyne detection in which the quantum fluctuations of the measured
field inevitably disturb the measurement and set the SNR of the output signal at the shot noise limit for a coherent input signal.

In what follows, let turn to the issue of negative CSD for squeezed signal. One should note that it is impossible in any way to experimentally distinguish the signal itself as given by Eq. (\ref{eq:sighetnn}) from the
corresponding CSD of Eq. (\ref{eq:csdhetnnnn}) in practical measurements. Therefore, from the detector one should have an output signal mixed with the CSD as follows,
\begin{equation}\label{eq:sigcsdhet}
\bar{S}_{\rm meas}=4^{-1}(\eta e c \varepsilon_0\varepsilon_{\rm l}\beta_{\rm s})^2e^{-2r}+2^{-1} c \varepsilon_0 \eta^2 e^2 \varepsilon_{\rm l}^2\hbar\omega_{\rm l}\left({\rm e}^{-2r}-1\right)B,\
\end{equation}
in which $B$ is the signal measurement bandwidth. Although the CSD, i.e., the second term on the right hand side of Eq. (\ref{eq:sigcsdhet}) is negative, it describes some quantum fluctuations that disturb the detected signal
in a way similar to usual quantum noises. This point will become clear if one pays attention to the quantum noise of a conventional detector that senses a squeezed optical signal, \cite{xie2021}
\begin{equation}\label{eq:pcshet}
\chi(\omega)=2 c \varepsilon_0 \eta^2 e^2 \varepsilon_{\rm l}^2\hbar\omega_{\rm l}+2 c \varepsilon_0 \eta^2 e^2 \varepsilon_{\rm l}^2\hbar\omega_{\rm l}\left({\rm e}^{-2r}-1\right),\
\end{equation}
where the first term on the right hand side denotes the shot noise of light and the second term describes additional quantum fluctuations introduced by optical squeezing. The value of second term is also negative and its
physical meaning is as follows: it represents some squeezing-induced quantum fluctuations opposite to the shot noise (vacuum fluctuations) and the two kinds of quantum fluctuations cancel each other leading to a total quantum
noise level below shot noise. Larger degree of squeezing indicates larger absolute values for the second term of (\ref{eq:pcshet}) and more quantum fluctuations introduced by optical squeezing. In the extreme case of
$r\rightarrow \infty$, squeezing introduced quantum fluctuations are maximal and cancel all vacuum fluctuations (shot noise), resulting in zero quantum noise as shown by Eq. (\ref{eq:pcshet}) for a regular detector.

Since the second term in Eq. (\ref{eq:pcshet}) is very similar to the CSD term in Eq. (\ref{eq:sigcsdhet}), they should describe similar noise origins in heterodyne detection. Therefore, a negative CSD does not mean lower
quantum noise in the detection; on the contrary, negative CSD's with larger absolute values correspond to  higher quantum noises into the detection. Therefore, if there are no classical noises, optical signals in squeezed
states will have poorer output SNR's in cross-correlatin heterodyne detection than those of coherent input signals.

But the real situations are more complex since there are always classical noises accompanied with any measurements. Let $N_{\rm cl}$ denotes the residual classical noises in the cross-correlation heterodyne detection, then the
detector will provide an output signal,
\begin{eqnarray}\label{eq:sigcsdnoisehet}
\bar{S}_{\rm meas}&=&4^{-1}(\eta e c \varepsilon_0\varepsilon_{\rm l}\beta_{\rm s})^2e^{-2r}+\nonumber\\
&&2^{-1} c \varepsilon_0 \eta^2 e^2 \varepsilon_{\rm l}^2\hbar\omega_{\rm l}\left({\rm e}^{-2r}-1\right)B + N_{\rm cl}.\
\end{eqnarray}
Now the sum of the total noises expressed as the last two terms on the right hand side of the above equation could be positive, negative, or zero depending on the value of each term. As an experimentally controllable
parameter, the value of $r$ in Eq. (\ref{eq:sigcsdnoisehet}) may be tuned to minimize the level of the total noises,
\begin{eqnarray}\label{eq:noisehet}
2^{-1} c \varepsilon_0 \eta^2 e^2 \varepsilon_{\rm l}^2\hbar\omega_{\rm l}\left({\rm e}^{-2r}-1\right)B + N_{\rm cl}=0,\
\end{eqnarray}
provided that the classical noises satisfy
\begin{eqnarray}\label{eq:clnoise}
 N_{\rm cl}<2^{-1} c \varepsilon_0 \eta^2 e^2 \varepsilon_{\rm l}^2\hbar\omega_{\rm l}B.\
\end{eqnarray}
Solving Eq. (\ref{eq:noisehet}) for $r$ gives
\begin{eqnarray}\label{eq:optr}
r=-2^{-1}\ln{\left[1-\frac{2N_{\rm cl}}{c \varepsilon_0 \eta^2 e^2 \varepsilon_{\rm l}^2\hbar\omega_{\rm l}B}\right]}.\
\end{eqnarray}
From Eqs. (\ref{eq:noisehet})-(\ref{eq:optr}) it follows that optical squeezing may be useful in the SNR improvement of the output signal of a cross-correlation heterodyne detector, when the level of classical noises meets the
requirment of inequality (\ref{eq:clnoise}), and there is an optimized degree of squeezing for most SNR enhancement as determined by Eq. (\ref{eq:optr}). Surely, the sum of classical noises may be frequency dependent, i.e.,
$N_{\rm cl}$ is a function of frequency. Then the optimized value of $r$ will be also frequency dependent for best SNR enhancement in cross-correlation heterodyne detection. This case is very different from that of traditional
heterodyne detection where the higher degree of squeezing the better SNR enhancement \cite{feng2016}.

\section{Conclusions}
We have developed a theory to describe the quantum behavior of cross-correlation heterodyne detectors and proven in theory their potential in suppression of the detection quantum noise below the shot noise level for output
signal SNR improvement. By calculating the cross spectral density of the heterodyne photocurrent fluctuations, we have shown that the noise performance of a cross-correlation heterodyne detector can exceed that of a regular
heterodyne detector and produce a sub-shot-noise output signal even if the input optical signal is in a coherent state. However, for detection of squeezed light the case becomes more complex. We have shown that the CSD of the
output signal will be negative in detection of squeezed light and discussed the physical meaning of negative CSD. And we have paid attention to how the degree of squeezing affects the output SNR of a cross-correlation
heterodyne detector, whose performance may be optimized by appropriately choosing the value of the degree of squeezing. This work should be of great interest for precision measurements in space-based gravitational wave
searching and other scientific research activities.

\textbf{Funding} This work was supported by the National Natural Science Foundation of China (12074110).

\textbf{Disclosures} The authors declare no conflicts of interest.

\textbf{Data availability} Data underlying the results presented in this paper are not publicly available at this time but may be obtained from the authors upon reasonable request.




\begin{thebibliography}{1}

\bibitem{armano2006}
M. Armano, et al., ``Sub-femto-g free fall for space-based gravitational wave observatories: LISA pathfinder results," Phys. Rev. Lett. \textbf{116}, 231101 (2006).

\bibitem{luo2016}
J. Luo, et al., ``Tianqin: a space-borne gravitational wave detector," Class. Quantum Gravity \textbf{33}, 035010 (2016).

\bibitem{babak2017}
S. Babak, J. Gair, A. Sesana, E. Barausse, C. F. Sopuerta,  C. P. L. Berry, E. Berti, P. Amaro-Seoane, A. Petiteau, and A. Klein, ``Science with the space-based interferometer LISA. V. Extreme mass-ratio inspirals," Phys. Rev.
D \textbf{95}, 103012 (2017).

\bibitem{zavattini2006}
E. Zavattini, et al., ``Experimental observation of optical rotation generated in vacuum by a magnetic field," Phys. Rev. Lett. \textbf{96}, 110406 (2006).

\bibitem{Sesana2016}
A. Sesana, ``Prospects for multiband gravitational-wave astronomy after GW150914," Phys. Rev. Lett. \textbf{116}, 231102 (2016).

\bibitem{xie2018}
B. Y. Xie and S. Feng, ``Squeezed-enhanced heterodyne detection of 10 Hz atto-Watt optical signals," Opt. Lett. \textbf{43}, 6073--6076 (2018).

\bibitem{michael2018}
E. A. Michael and F. E. Besser, ``On the Possibility of Breaking the Heterodyne Detection Quantum Noise Limit with Cross-Correlation," IEEE Access \textbf{6}, 45299--45316 (2018).

\bibitem{mitsui2013}
T. Mitsui and K. Aoki, ``Measurements of liquid surface fluctuations at sub-shot-noise levels with Michelson interferometry," Phys. Rev. E \textbf{87}, 042403 (2013).

\bibitem{martynov2017}
D. V. Martynov, et al., ``Quantum correlation measurements in interferometric gravitational-wave detectors," Phys. Rev. A \textbf{95}, 043831 (2017).

\bibitem{venneberg2022}
J. R. Venneberg and B. Willke, ``Quantum correlation measurement of laser power noise below shot noise," Opt. Continuum \textbf{1}, 1077--1084 (2022).

\bibitem{glauber1963}
R. J. Glauber, ``Coherent and incoherent states of the radiation field," Phys. Rev. \textbf{131}, 2766--2788 (1963).

\bibitem{mandel1995}
L. Mandel and E. Wolf, {\it Optical Coherence and Quantum Optics} (Cambridge University, 1995).

\bibitem{feng2016}
S. Feng, D. C. He and B. Y. Xie, ``Quantum theory of phase-sensitive heterodyne detection," J. Opt. Soc. Am. B \textbf{33}, 1365--1372 (2016).

\bibitem{ou1987}
Z. Y. Ou, C. K. Hong and L. Mandel, ``Coherent properties of squeezed light and the degree of squeezing," J. Opt. Soc. Am. B \textbf{4}, 1574--1587 (1987).

\bibitem{Slusher1985}
R. E. Slusher, L. W. Hollberg, B. Yurke, J. C. Mertz and J. F. Valley, ``Observation of squeezed states generated by four-wave mixing in an optical cavtiy," Phys. Rev. Lett. \textbf{55}, 2409--2412 (1985).

\bibitem{Wu1986}
L.-A. Wu, H. J. Kimble, J. L. Hall and H. F. Wu, ``Generation of squeezed states by Parametric Down-Conversion," Phys. Rev. Lett. \textbf{57}, 2520--2523 (1986).

\bibitem{caves1982}
C. M. Caves, ``Quantum limits on noise linear amplifiers," Phys. Rev. D \textbf{26}, 1817--1839 (1982).

\bibitem{xie2021}
B. Y. Xie and S. Feng, ``Heterodyne detection enhanced by quantum correlation," Chin. Opt. Lett. \textbf{19}, 702701 (2021).









\end{thebibliography}
\end{document}